\documentclass{article}
\usepackage{graphicx}
\usepackage[english]{babel}

\title{
\includegraphics[width=0.35\textwidth]{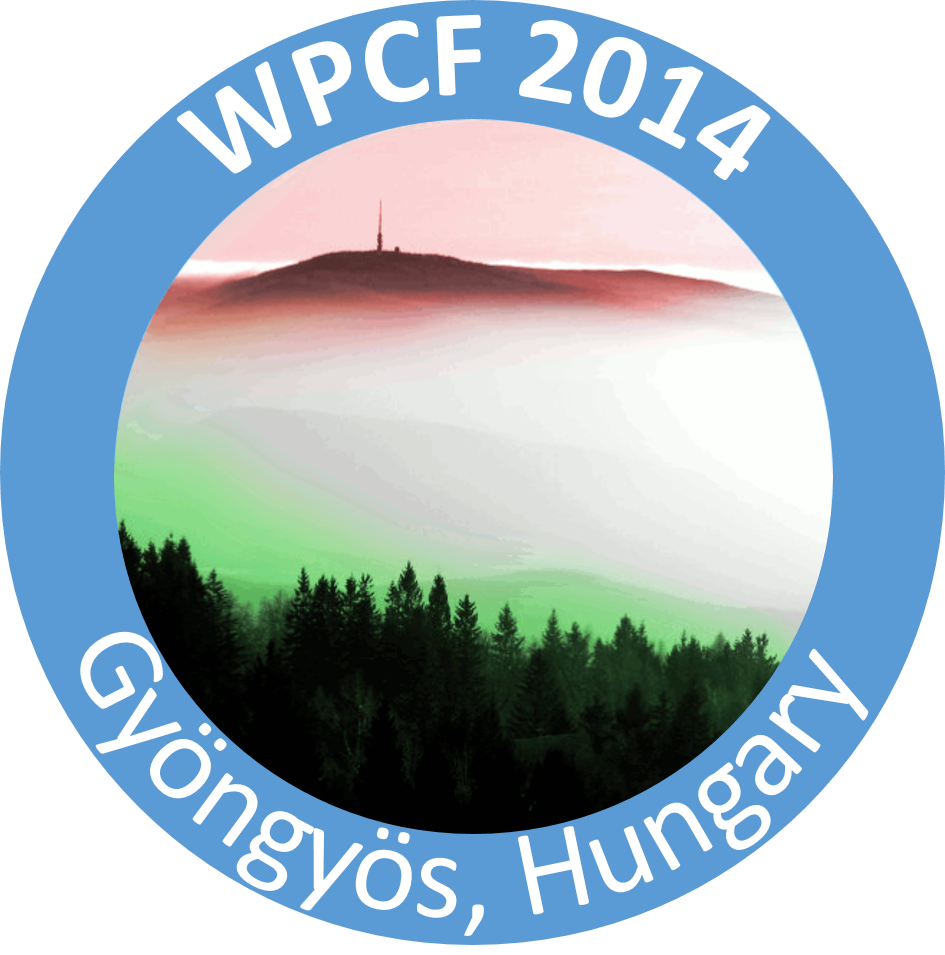}\\[1cm]
Chaoticity and Coherence in \\ Bose-Einstein Condensation and
Correlations\footnote{ Invited talk presented at the Xth Workshop on
  Particle Correlations and Femtoscopy \hspace*{0.5cm}(WPCF14) at
  Gy\"ongy\"os, Hungary on August 25 to 29, 2014 \protect\newline
\hspace*{0.3cm}$\dagger$ Speaker
}
}
\author{{Cheuk-Yin Wong$^1$$^\dagger$ 
,Wei-Ning Zhang$^2$, Jie Liu$^2$, Peng  Ru$^2$,}\\[1ex]
$^1$Physics Division,Oak Ridge National Laboratory,Oak Ridge,TN,USA \\
$^2$School of Physics and Optoelectric Technology,\\Dalian University of Technology, Dalian, Liaoning, China\\
}

\begin{document}

\def\bb    #1{\hbox{\boldmath${#1}$}}

\fontfamily{lmss}\selectfont
\maketitle
\begin{abstract}
We review the properties of chaoticity and coherence in Bose-Einstein
condensation and correlations, for a dense boson system in its
mean-field represented approximately by a harmonic oscillator
potential.  The order parameter and the nature of the phase transition
from the chaotic to the condensate states are studied for different
fixed numbers of bosons.  The two-particle correlation function in
momentum space is calculated to investigate how the Bose-Einstein
correlation depends on the degree of condensation and other momentum
variables.  We generalize the Bose-Einstein correlation analysis to
three-particle correlations to show its dependence on the degree of
condensation.
\end{abstract}
\vspace*{0.3cm}

\section{Introduction}

As is well known, a fundamental assumption for the occurrence of
Bose-Einstein correlation (BEC) is the presence of a chaotic source of
identical bosons \cite{Han56,Gla63}.  The Bose-Einstein correlation
occurs in a chaotic source but not in a coherent source
\cite{Gyu79,Vos94,Wie99,Wei00,Lis05,Won94}.

The properties of chaoticity and coherence are complementary
attributes.  Both chaoticity and coherence should be examined on equal
footings in a single theoretical framework with the description of
both the BE condensation and BE correlations.  In such a unified
framework, it is then possible to investigate not only the states of
chaoticity and coherence, but also the transition from a chaotic
state to a coherent state.  How can the degrees of chaoticity or
coherence be quantified?  Is the transition from a chaotic state to a
coherent state a first-order with a sudden onset, or is it a
gradual transition that is closer to a second-order?  What is the
relevant order parameter that best describes the transition?  How does
Bose-Einstein condensation  quantitatively affects the two-particle and
three-particle Bose-Einstein correlations?

Questions of Bose-Einstein correlations and condensation arise not only
in atomic physics \cite{Pol96,Nar99,Gom06} but also in high-energy
heavy-ion collisions \cite{Wie99,Wei00,Won94} where pions are the most
copiously produced particles.  The use of two-pion Bose-Einstein
correlations to probe the source coherence was proposed at the end of
1970s \cite{Fow77,Gyu79}.  The introduction of the ``chaoticity"
parameter $\lambda$ of BEC in pions is only a tool to represent
experimental data.  However, the experimental measurement of $\lambda$
is beset by the presence of many other effects such as particle
misidentification, long-live resonance decay, final state Coulomb
interaction, non-Gaussian source distribution,
etc. \cite{Wie99,Lis05}.  The explanation of the experimental
$\lambda$ results remains an open question.  In 1993, S. Pratt
proposed a pion laser model in high energy collisions and studied the
influence of pion laser on two-pion Bose-Einstein correlation function
and the chaoticity parameter \cite{Pra93}.  In 1998, T. Cs\"{o}rg\H{o}
and J. Zim\'{a}nyi investigated the effect of Bose-Einstein
condensation on two-pion Bose-Einstein correlations \cite{CsoZim98}.
They utilized Gaussian formulas describing the space and momentum
distributions of a static non-relativistic boson system, and
investigated the influence of the condensation on pion multiplicity
distribution.  In 2007, C. Y. Wong and W. N. Zhang studied how
$\lambda$ in Bose-Einstein correlations depends on the degree of
Bose-Einstein condensation or chaoticity, for static non-relativistic
and relativistic boson gases within a spherical mean-field harmonic
oscillator potential \cite{Won07}.  The model can be analytically
solved in the non-relativistic case and be used in atomic physics
\cite{Pol96,Nar99,Gom06}.  The limiting conditions and circumstances
under which the parameter $\lambda$ can be approximately related to
the degrees of chaoticity were clarified \cite{Won07}.  A similar
study for cylindrical static boson gas sources was completed
\cite{Liu13} and the chaoticity parameter $\lambda$ in two-pion
Bose-Einstein correlations in an expanding boson gas model was
recently examined \cite{LiuZha14}.  The investigation of chaoticity
and coherence was also carried out using a model of q-deformed
oscillator algebraic commutative relations \cite{Gav06} and the model
of partial indistinguishability and coherence of closely located
emitters \cite{Sin13}.  In another related topic, initial conditions
such as the color-glass condensate (CGC) with the coherent production
of partons \cite{McL94} in heavy-ion collisions may also lead to
condensate formation \cite{Bla12}.

Recently, experimental investigation of the source coherence in Pb-Pb
collisions at $\sqrt{s_{NN}}=2.76$ TeV at the Large Hadron Collider
(LHC) was carried out by the ALICE collaboration \cite{Abe14}.  A
substantial degree of source coherence was measured \cite{Abe14} using
a new three-pion Bose-Einstein correlations technique with an
improvement over past efforts \cite{NA44,WA98,STAR,Mor05}.  Earlier
work on three-particle correlations were carried out in
\cite{Wei00,Pra93,Zaj87,Biy90,And91,Zha93,Cha95,Hei97,Nak99,Cso02}.

A proper theoretical framework to study the above topics is the theory
of the Bose-Einstein condensation and correlations in their own mean
field potential \cite{Won07}.  We would like to review the essential
elements here and examine further the related question of three-body
correlations.

\section{Bose-Einstein Condensation for attractively  Interacting Bosons}

We seek a description of chaoticity in Bose-Einstein correlations
through the consideration of Bose-Einstein condensation.  Why is
Bose-Einstein condensation relevant to Bose-Einstein correlations
(BEC)?  Glauber in many private communications and in his talk in
QM2005 suggested that the consistent experimental observations of
$\lambda<$ 1 may be due partly to the coherence of the pions in Bose
Einstein correlations \cite{Gla06}.  Furthermore, there have been
major advances in Bose-Einstein condensation in atomic physics
\cite{Pol96,Nar99,Gom06}.
In particular, the works of Politzer \cite{Pol96}, and Naraschewski \&
Glauber \cite{Nar99} reveals that BE condensation and the BE
correlations are intimately related.

We envisage the possibility of the occurrence of a Bose-Einstein
condensation in dense boson media of identical bosons with the
following reasoning \cite{Won07,Liu13,LiuZha14}
\begin{enumerate}

\item
Identical bosons with mutual attractive interaction generate a mean
field potential, which depends on the boson density $\rho(r)$ as
\cite{Gla59}
\begin{eqnarray}
V(r) = - 4 \pi f(0) \rho (r) \sim \frac{1}{2} \hbar \omega \left ( \frac {r}{a}
 \right ) ^2,
\end{eqnarray}
where $f(0)$ is the forward scattering amplitude, and $a$ is the
length scale that defines the spatial region of boson occupation.

\item
Therefore, for a given length scale $a$, the
$\hbar \omega$ of the underlying mean-field potential increases with
increasing density $\rho$ of the produced bosons.

\item
The order parameter that determines the degree of
BE coherence or chaoticity is $T/\hbar \omega$.  Thus the order
parameter $T/\hbar \omega$ decreases with increasing boson density.

\item
For a given temperature $T$ at freeze out, a high density of produced
bosons will lead to a lower value of the order parameter $T/\hbar
\omega$, which in turn will lead to a greater condensate fraction
$f_0$=$N_0/N$, where $N$ is the total number of bosons and $N_0$ is
the number of bosons in the lowest state.  A greater condensate
fraction $f_0$ brings about a greater coherence in Bose-Einstein
correlations and a reduction in the degree of chaoticity.

\end{enumerate}

In high energy heavy-ion collisions when bosons (gluons or pions) are
copiously produced within a small region in a short time interval, the
density of the bosons increases as the collision energy increases.
Following the above reasoning, generalized to systems with
differential transverse and longitudinal spatial distributions, we
expect the occurrence of boson condensation in high energy heavy-ion
collisions at some high collision energies.  It is useful to examine
the Bose-Einstein condensation for bosons in an exactly solvable
model.

\section{Bose-Einstein Condensation for Bosons in a Spherical Harmonic Oscillator Potential}

We consider a system of $N$ bosons in a spherical harmonic oscillator
potential, which arises either externally or from the bosons' own
mean-fields.  We study how the occupation numbers of different states
change as a function of the temperature $T$, in relation to the
oscillator frequency $\hbar \omega$.  Bose-Einstein condensation
occurs when the occupation number $N_0$ for the lowest state (the
condensate state) is a substantial fraction of the total particle
number $N$.  The degree of coherence or chaoticity is quantified by
the condensate fraction $f_0=N_0/N$, which varies as a function of the
order parameter $T/\hbar \omega$.

In such a study, it is important to use the proper statistical
ensemble \cite{Pol96}.  In a grand canonical ensemble, we fix the
chemical potential $\mu$ and the temperature $T$, and we allow the
number of particles $N_n$ in the $n$-th single-particle state to vary.
We obtain the average occupation number for the single-particle state
$n$ to be $N_n=\langle a_n^+ a_n\rangle$.  The square fluctuation of
$N_n$ is then given by
\begin{eqnarray}
\langle (a_n^+ a_n  - \langle a_n^+ a_n\rangle )^2 \rangle \approx N_n (N_n +1).
\end{eqnarray}

As the fluctuation of $N_n$ in a grand canonical ensemble is of the
same order as the occupation number itself, we cannot treat the lowest
$n=0$ state in the grand canonical ensemble.  The lowest $n=0$ state
needs to be treated in the canonical ensemble with a fixed total
number of bosons.

It was shown however that while the lowest $n=0$ state needs to be
treated in the canonical ensemble, the $n > 0$ state can be treated in
the grand canonical ensemble without incurring large errors
\cite{Pol96}.  We shall follow such a description for the ensemble of
$N$ identical bosons in a spherical harmonic oscillator potential.  In
such a canonical ensemble for the lowest $n$=0 state but a grand
canonical ensemble for the $n$$ >$0 states, the total number of bosons
is fixed and yields the condensate number condition
\begin{eqnarray}
N=N_0+\!\!\!\! \sum_{n=1,2,3,...}\!\!\!\! \!\!N_n=\frac{z}{1-z}  + \sum_{n=1,2,3,...} \frac{g_n z e^{-( \epsilon_n-\epsilon_0) /T}}
{1-z e^{-( \epsilon_n-\epsilon_0)/T}},
\label{eq6}
\end{eqnarray}
where $z=e^{\mu/T}$ is the fugacity of the system, $g_n$ is the
degeneracy number $g_n$=$(n$+1)$(n$+2)/2 for the $n$-th
single-particle level, and $\epsilon_n$ is the single-particle energy
in the spherical harmonic oscillator potential
\begin{eqnarray}
\epsilon_n = (n+3/2) \hbar \omega.
\end{eqnarray}
For a given $N$, equation (\ref{eq6}) contains only a single unknown,
$z$, which can be solved as a function of the order parameter $T/\hbar
\omega$.  The solutions of $z$ for $N$=25, 500, 1000 and 2000 are
given in Fig.\ 1, and the corresponding condensate fractions
$f_0=N_0/N$ are shown in Fig. 2.
\begin{figure}[h]
\hspace*{2.0cm}
\includegraphics[scale=0.5]{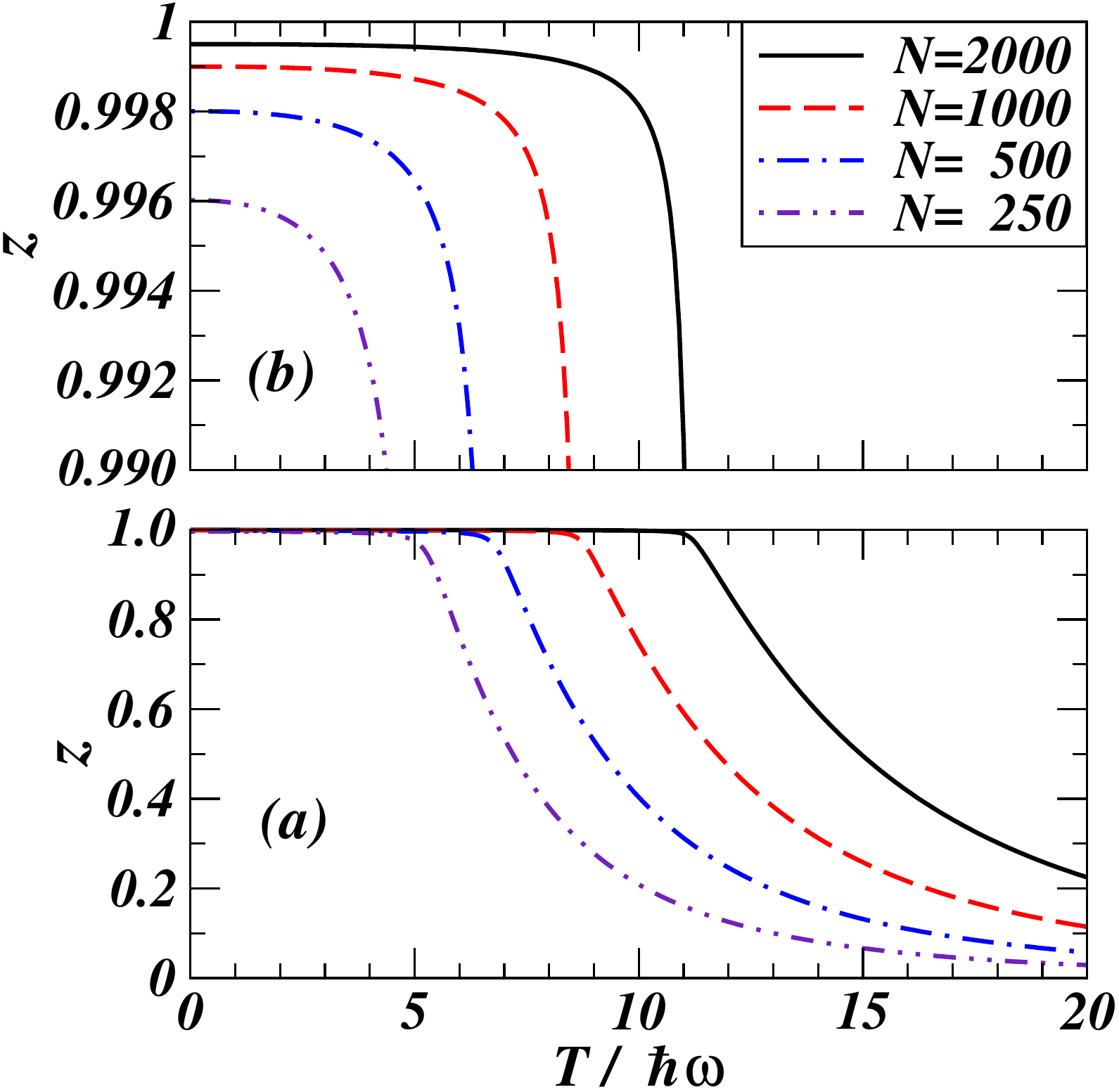}
\caption{(Color online) (a) The fugacity parameter $z$ satisfying the
  condensate number condition Eq.\ (\ref{eq6}) for different boson
  numbers $N$ in a spherical harmonic oscillator potential, as a
  function of the order parameter $T/\hbar \omega$ and ($b$) an
  expanded view in the $z\sim 1$ region.  }
\end{figure}

 We observe in Fig. 1 that the fugacity parameter $z$ is close to
 unity in the strongly coherent region at low temperatures.  In fact,
 the fugacity parameter $z$ at $T$=0 assumes the value $ z(T$=$0)=
 {N}/(N+1).  $ For a given boson number $N$, the fugacity $z$
 decreases very slowly in the form of a plateau, as the temperature
 increases from $T=0$.  The plateau region persists until the
 condensate temperature $T_c$ is reached, and $z$ then decreases
 rapidly thereafter.  The greater the number of bosons $N$, the
 greater is the plateau region, as shown in Fig.\ 1($b$).  For
 example, for $N=2000$ the value of $z$ is close to unity for
 $0<T/\hbar \omega<11$ in the plateau,.

\begin{figure}[h]
\hspace*{0.cm}
\includegraphics[scale=0.65]{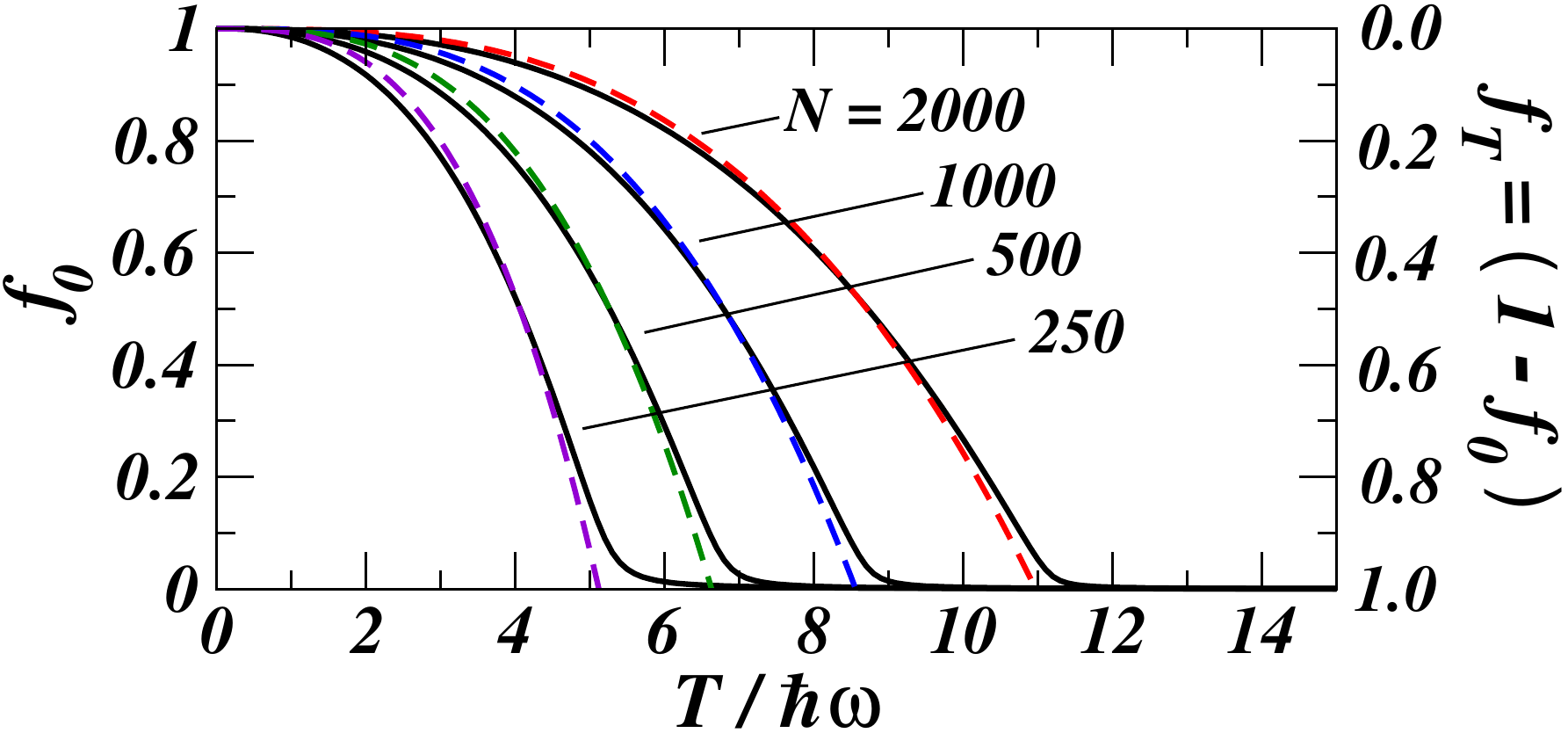}
\caption{(Color online) Solid curves represent the condensate
  fractions $f_0(T)$, calculated with the condensate number condition
  Eq.\ (\ref{eq6}), as a function of $T/\hbar \omega$ for different
  boson numbers $N$ in a spherical harmonic oscillator potential.  The
  abscissa labels for the corresponding chaotic fraction
  $f_T(T)$=$[1-f_0(T)]$ are indicated on the right.  The dashed curves
  are the fits to the solid curve results of $f_0(T)$ with the
  function 1$-[(T/\hbar \omega)/(T_c/\hbar \omega)]^3$ of
  Eq.\ (\ref{fit}) where the values of $T_c/\hbar \omega$ for
  different $N$ values are listed in Table I.  }
\end{figure}

We note in Fig. 2 that for a given value of the total number of bosons
$N$ in the spherical harmonic oscillator potential, the condensate
fraction $f_0$ is close to unity when the order parameter $T/\hbar
\omega$ is below a limit, and this limit depends on $N$.  We can plot
the condensate fraction $f_0$ as a function of the order parameter
$T/\hbar \omega$.  The functional form of $f_0(T)$ can be approximated
by
\begin{eqnarray}
\label{fit}
f_0(T)=
\begin{cases} 
{
   1-[(T/\hbar \omega) /(T_c/\hbar \omega)]^3  & for $(T/\hbar \omega) \le (T_c/\hbar \omega),$   \cr
  O(1/N) \to 0  &   for $(T/\hbar \omega) \ge (T_c/\hbar \omega).$  \cr
}
\end{cases}
\end{eqnarray}
The results from the above one-parameter fit to $f_0(T)$ are shown as
the dashed curves in Fig.\ 2, to be compared with the $f_0(T)$
calculated with the condensate configuration condition
Eq.\ (\ref{eq6}) shown as the solid curves.  The values of $T_c/\hbar
\omega$ that give the best fit to $f_0(T)$ for different $N$ values
are listed in Table I.

The above results provide a comprehensive description for the
transition from a chaotic state to a coherent state.  Fig.\ 2
indicates that the transition from the completely chaotic state with
$f_0$=0 to the state of coherence with $f_0 $$\to$1 is a
second-order-type transition under a gradual decrease of the order
parameter $T/\hbar \omega$.  It is not a first-order phase transition.

\vskip 0.4cm \centerline{Table I.  Critical order parameter $T_c/\hbar
  \omega$ obtained from (i) fitting $f_0$ as a function } \centerline{
  of $T/\hbar \omega$ with Eq. (\ref{fit}), and from (ii) the
  analytical formula of Eq.\ (\ref{Tc}). }  {\vskip 0.3cm\hskip 0.2cm
\begin{tabular}{|c|c|c|}
\hline
{\rm Number of Bosons }& ~~~$T_c/\hbar \omega$ obtained ~~&  $T_c/\hbar \omega$ obtained
                 \\
         $N$        & from fitting $f_0$ with Eq. (\ref{fit})   &  with Eq.(\ref{Tc})
                   \\
\hline
       2000      &  10.97 & 11.00
\\ \cline{1-3}  
       1000      &   8.56 & 8.53
\\ \cline{1-3}  
        500      &   6.63 & 6.62
\\ \cline{1-3}  
        250      &   5.12 & 5.13
\\ \cline{1-3}  
\end{tabular}
}
\vspace*{0.3cm}

It is remarkable that the critical order parameter $T_c/\hbar \omega$
and the boson number $N$ obeys the following simple relationship
\begin{eqnarray}
T_c/\hbar \omega &=& 0.6777 N^{0.36666},
\label{Tc}
\end{eqnarray}
as shown by the third column in Table I.  Thus, the knowledge of $N$
suffices to determine the critical order parameter $T_c/\hbar \omega$
by the above simple equation and the knowledge of $T_c/\hbar \omega$
subsequently yields the approximate condensate fraction at all other
temperatures by Eq.\ (\ref{fit}).

\section{Single-particle  and Two-Particle Density \\Matrices in Momentum Space}

The determination of the fugacity $z$ from the condensate number
condition (\ref{eq6}) allows the calculation of various physical
quantities.  Specifically, the one-body density matrix in momentum
space is given by
\begin{eqnarray}
\label{one}
G^{(1)}(\bb{p}_1,\bb{p}_1')= \sum_{n=0}^{\infty} 
u_n^*(\bb{p}_1')
u_n  (\bb{p}_1) \langle \hat a_n^\dagger \hat a_n \rangle,
\end{eqnarray}
where $u_n(\bb p)$ is the single-particle wave function and the
occupation number\break  $N_n$=$\langle \hat a_n^\dagger \hat a_n \rangle$
can be inferred from the terms in the summation in Eq. (\ref{eq6}).
The two-particle density matrix in momentum space
\begin{eqnarray}
G^{(2)}(\bb{p}_1,\bb{p}_2; \bb{p}_1',\bb{p}_2') 
=
\sum_{\rm klmn}
u_{\rm k}^*(\bb{p}_1') u_{\rm l}^*(\bb{p}_2')
u_{\rm m}(\bb{p}_2) u_n (\bb{p}_1)
 \langle \hat a_{\rm k}^\dagger \hat a_{\rm l}^\dagger 
\hat a_{\rm m} \hat a_{\rm n}
\rangle
\end{eqnarray}
can be written
in terms of one-body density matrices as \cite{Nar99,Won07}
\begin{eqnarray}
\label{eq10}
\!\!\!\!\!G^{(2)}(\bb{p}_1,\bb{p}_2; \bb{p}_1',\bb{p}_2') 
=G^{(1)}(\bb{p}_1,\bb{p}_1') G^{(1)}(\bb{p}_2,\bb{p}_2')
+ G^{(1)}(\bb{p}_1,\bb{p}_2')G^{(1)}(\bb{p}_2,\bb{p}_1')
\nonumber\\
~~~~~~+\sum_{ n=0}^{\infty}
u_{ n}^*(\bb{p}_1') u_{ n}^*(\bb{p}_2')
u_{n}(\bb{p}_2) u_n (\bb{p}_1)
\biggl  \{ \langle 
\hat a_n^\dagger \hat a_n \hat a_n
\rangle
- 2  
\langle \hat a_n^\dagger \hat a_n
\rangle
\langle \hat a_n^\dagger \hat a_n
\rangle \biggr \}.
\label{eq9}
\end{eqnarray}
\begin{figure}[h]
\hspace*{1.0cm}
\includegraphics[scale=0.70]{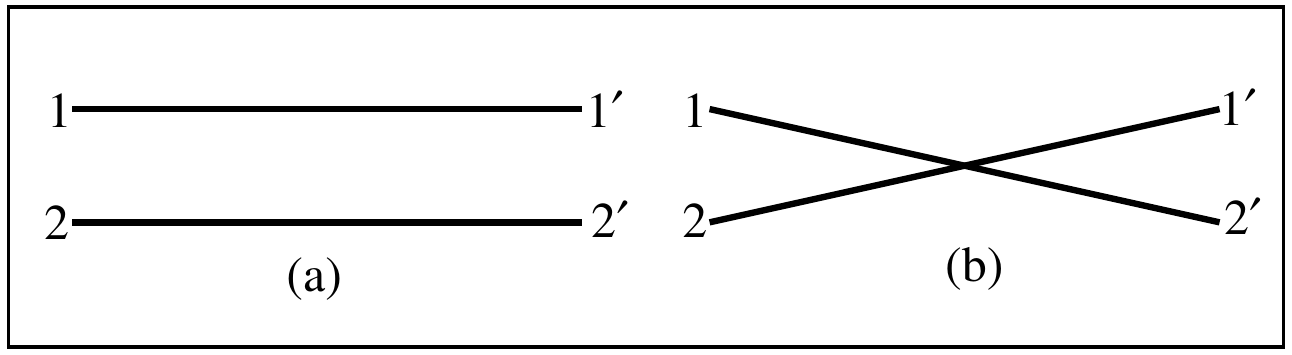}
\caption{Two-particle distribution function expanded in terms of
  products of one-particle distribution functions in uncorrelated
  mean-field approximation. }
\end{figure}

The uncorrelated part in the first two terms of the above two-particle
density matrix, $ G^{(1)}(\bb{p}_1,\bb{p}_1')
G^{(1)}(\bb{p}_2,\bb{p}_2') +
G^{(1)}(\bb{p}_1,\bb{p}_2')G^{(1)}(\bb{p}_2,\bb{p}_1')$, is
represented schematically by the diagram in Fig.\ 3.  Our task is to
obtain the correlated part arising from Bose-Einstein condensation
represented by the last term in Eq.\ (\ref{eq9}).

In the limit of a large number of bosons $N$ in a grand canonical
ensemble for the non-condensed states, the contributions from the set
of $\{{ n}>0\}$ states in the summation in Eq.\ (\ref{eq10}) can be
neglected. We are left with only the ${ n}=0$ condensate state
contribution for this summation.

To describe the contribution from the $n=0$ condensate state, we shall
follow Ref.\ \cite{Pol96,Nar99} and use the canonical ensemble which
gives the canonical fluctuation
\begin{eqnarray}
\langle ( \hat a_{\rm 0}^\dagger \hat a_{\rm 0}
   -\langle\hat a_{\rm 0}^\dagger \hat a_{\rm 0}\rangle )^2 \rangle 
= \langle \hat a_0^\dagger \hat a_0^\dagger \hat a_0 \hat a_0
\rangle- \langle \hat a_0^\dagger \hat a_0\rangle  
  \langle \hat a_0^\dagger \hat a_0\rangle
=O(N_0).
\end{eqnarray}
Thus, we have
\begin{eqnarray}
\langle \hat a_{\rm 0}^\dagger 
\hat a_{\rm 0}^\dagger \hat a_{\rm 0} \hat a_{\rm 0}
\rangle
- 2  
\langle \hat a_{\rm 0}^\dagger \hat a_{\rm 0}
\rangle
\langle \hat a_{\rm 0}^\dagger \hat a_{\rm 0}
\rangle 
= - \langle \hat a_0^\dagger \hat a_0\rangle  
  \langle \hat a_0^\dagger \hat a_0\rangle + O(N_0).
\end{eqnarray}
In the limit of a large number of particles, we can neglect the last
term $O(N_0)$ in the above equation which is small in comparison with
the first term of order $N_0^2$.  The two-particle distribution of
Eq.\ (\ref{eq10}) is therefore
\begin{eqnarray}
\label{eq18}
\hspace*{-0.8cm}
G^{(2)}(\bb{p}_1,\bb{p}_2; \bb{p}_1,\bb{p}_2) 
=G^{(1)}(\bb{p}_1,\bb{p}_1) G^{(1)}(\bb{p}_2,\bb{p}_2)
+ | G^{(1)}(\bb{p}_1,\bb{p}_2)|^2~~~~~~~~~
\nonumber\\
 -N_0^2 
|u_0(\bb{p}_1)|^2 | u_0(\bb{p}_2)|^2, 
\end{eqnarray}
which gives the conditional probability for the occurrence of a pion
of momentum $\bb{p}_1$ in coincidence with another identical pion of
momentum $\bb{p}_2$.

\section{Two-Particle Momentum Correlation Function}

In BE correlation  measurements, we normalize the probability
relative to the probability of detecting particle $\bb{p}_1$ and
$\bb{p}_2$, and define the momentum correlation function
$C(\bb{p}_1,\bb{p}_2)$ as
\begin{eqnarray}
C(\bb{p}_1,\bb{p}_2)=
\frac {G^{(2)}(\bb{p}_1,\bb{p}_2; \bb{p}_1,\bb{p}_2)}
      {G^{(1)}(\bb{p}_1,\bb{p}_1) G^{(1)}(\bb{p}_2,\bb{p}_2)}.
\end{eqnarray}
It is convenient to introduce the average and the relative momenta 
of the pair
\begin{eqnarray}
\bb{p}=(\bb{p}_1+\bb{p}_2)/2,
~~~~~~~~\bb{q}=\bb{p}_1-\bb{p}_2.
\end{eqnarray}
The momentum correlation function can be expressed as a function of
the kinematic variables $\bb p_1$ and $\bb p_2$ or alternatively of
$\bb{p}$ and $\bb{q}$.  From Eq. (\ref{eq18}), we have the general
expression for the correlation function
\begin{eqnarray}
\label{eq22}
C(\bb{p},\bb{q})=C(\bb{p}_1,\bb{p}_2)=1+
\frac {|G^{(1)}(\bb{p}_1,\bb{p}_2)|^2-N_0^2 |u_0(\bb{p}_1)|^2
                                                |u_0(\bb{p}_2)|^2}
      {G^{(1)}(\bb{p}_1,\bb{p}_1) G^{(1)}(\bb{p}_2,\bb{p}_2)}.
\end{eqnarray}
This is the general Bose-Einstein correlation function for all
situations: coherent, chaotic, and the transition between coherent and
chaotic systems.

The evaluation of the correlation function $C({\bb p}, {\bb q})$ in
Eq.\ (\ref{eq22}) requires the knowledge of $G^{(1)}(\bb p_1,\bb p_2)$
and the ground state wave function $u_0(\bb p_1)$.  For a system of
bosons in a spherical harmonic oscillator, the wave functions are all
known, and the correlation function can be written out analytically.
Specifically, we have
 
\begin{eqnarray}
&&\hspace*{-1.2cm}G^{(1)}(\bb{p}_1,\bb{p}_2) 
=\sum_{k=1}^\infty z^k 
{\tilde G}_0 (\bb{p}_1,\bb{p}_2; k \beta \hbar \omega),
\label{eq23}\\
&&\hspace*{-1.2cm}{\tilde G}_0 (\bb{p}_1,\bb{p}_2; \tau)
\!\!= \!\!\left (  \!\!\frac {a^2}{\pi \hbar^2 (1-e^{-2\tau})}\! \right )^{\! \! \!3/2}
\!\!\!\!\exp \!\left (\!\! -\frac {a^2}{\hbar^2} 
\frac {(\bb{p}_1^2 \!+\! \bb{p}_2^2)(\cosh\tau \!-\!1)\!+\!(\bb{p}_1\!-\!\bb{p}_2)^2}
      {2\sinh \tau} \! \right )\!\!, 
\end{eqnarray}
and the ground state wave function is
\begin{eqnarray}
u_0(\bb {p})=\left (\frac{a^2}{\pi \hbar^2}\right )^{3/4} 
\exp \left \{ -\frac{a^2}{\hbar^2}\frac{\bb {p}^2}{2} \right \}.
\label{eq23A}
\end{eqnarray}
The knowledge of the single-particle $G^{(1)}(\bb{p}_1,\bb{p}_2)$ and
$u_0(\bb p)$ will then allow the determination of the two-particle
correlation function $C(\bb p, \bb q)$.

The correlation function $C(\bb p, \bb q)$ in Eq. (\ref{eq22})
possesses the proper coherent and chaotic limits.  For a nearly
completely coherent source with almost all particles populating the
ground condensate state, $N_0 \to N$, the two terms in the numerator
cancel each other and we have $C(\bb p,\bb q)=1$, with the absence of
the BE correlation.  For the other extreme of a completely chaotic
source with $N_0$$\ll$$N$, the second term in the numerator
proportional to $N_0^2$ in Eq. (\ref{eq22}) gives negligible
contribution and can be neglected.  The correlation function $C(\bb
p,\bb q)$ then becomes the usual BE correlation for a completely
chaotic source.

\section{Evaluation of the Two-Particle
Momentum Correlation Function}

\begin{figure}[h]
\hspace*{2.0cm}
\includegraphics[scale=0.45]{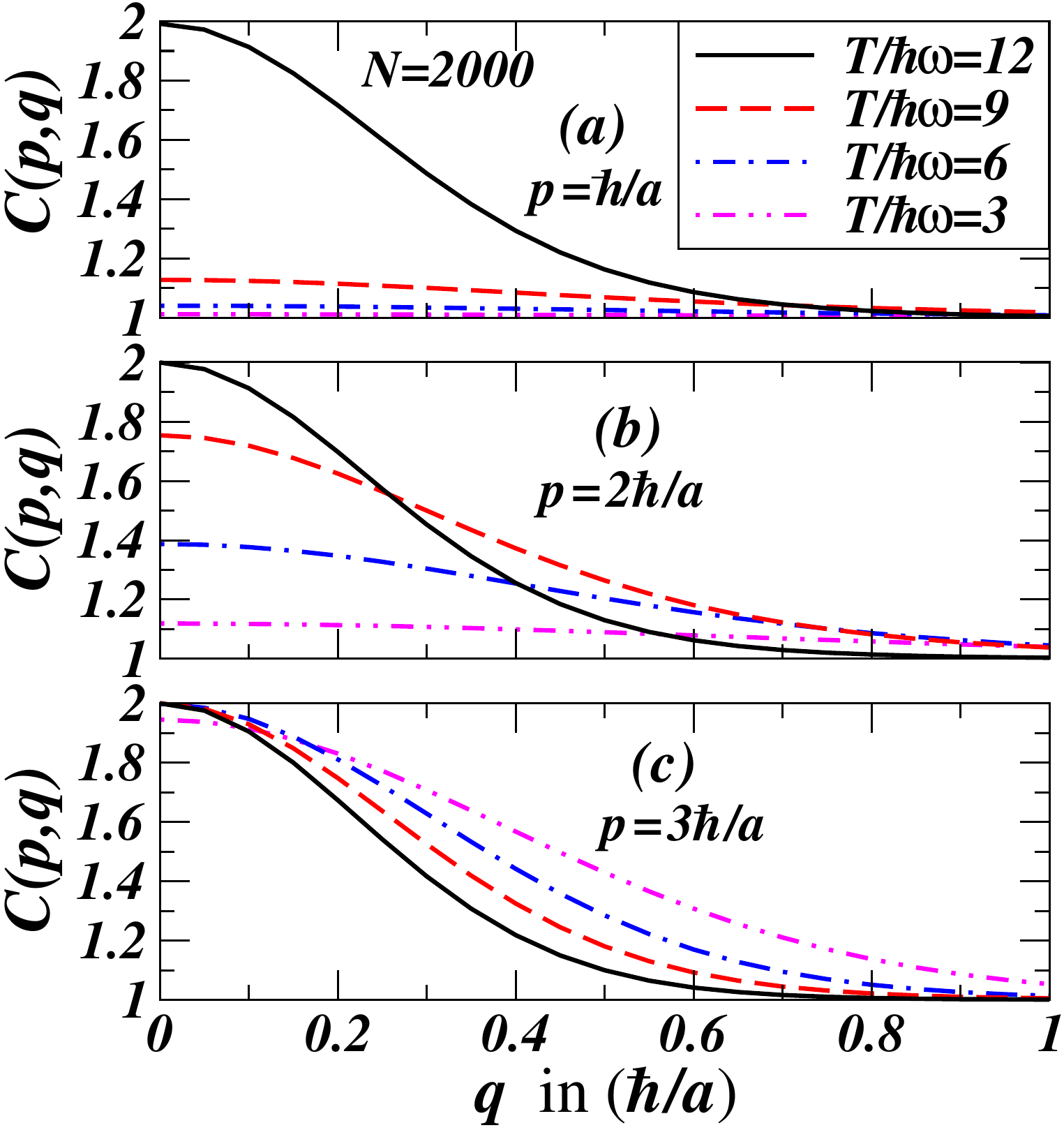}
\caption{(Color online) The correlation function $C(\bb p, \bb q)$ at
  different values of the pair average momentum $pa/\hbar$ and
  temperatures.  Figures (a), (b), and (c) are for $p$=1, 2, and
  3$\hbar/a$, respectively.  }
\end{figure}

For a given number of bosons $N$ in a spherical harmonic oscillator,
the solution of fugacity $z$ obtained as a function of the order
parameter $T/\hbar \omega$ allows us to evaluate the momentum
correlation function $C(p,q)$ with Eqs.\ (\ref{eq22})-(\ref{eq23A}).
In Fig.\ 4, we show $C(p,q)$ for example for the case of $N$=$2000$
for which the critical order parameter is $T_c/\hbar \omega$=10.97, as
tabulated in Table I.  We observe that the correlation function is a
complicated function of the average pair momentum $p$ and the order
parameter $T/\hbar \omega$.  For $p$=$\hbar/a$ in Fig. 4($a$), the
correlation function $C(p,q)$ at $q$=0 is close to unity for
temperatures below and up to $T/\hbar \omega$=9 (below $ T_c/\hbar
\omega$), but increases to 2 rather abruptly at $ T/\hbar \omega$=12,
(above $T_c/\hbar \omega$).  For $p$=2$\hbar/a$ in Fig. 3($b$), the
correlation function $C(p,q)$ at $q=0$ is substantially above unity
and increases gradually as temperature increases.  For $p$=$3\hbar/a$
in Fig. 4($c$), the correlation function $C(p,q)$ at $q$=0 is about 2
for all cases of temperatures examined.
\begin{figure}[h]
\vspace*{-2.3cm}
\hspace*{-2.1cm}
\includegraphics[scale=0.65]{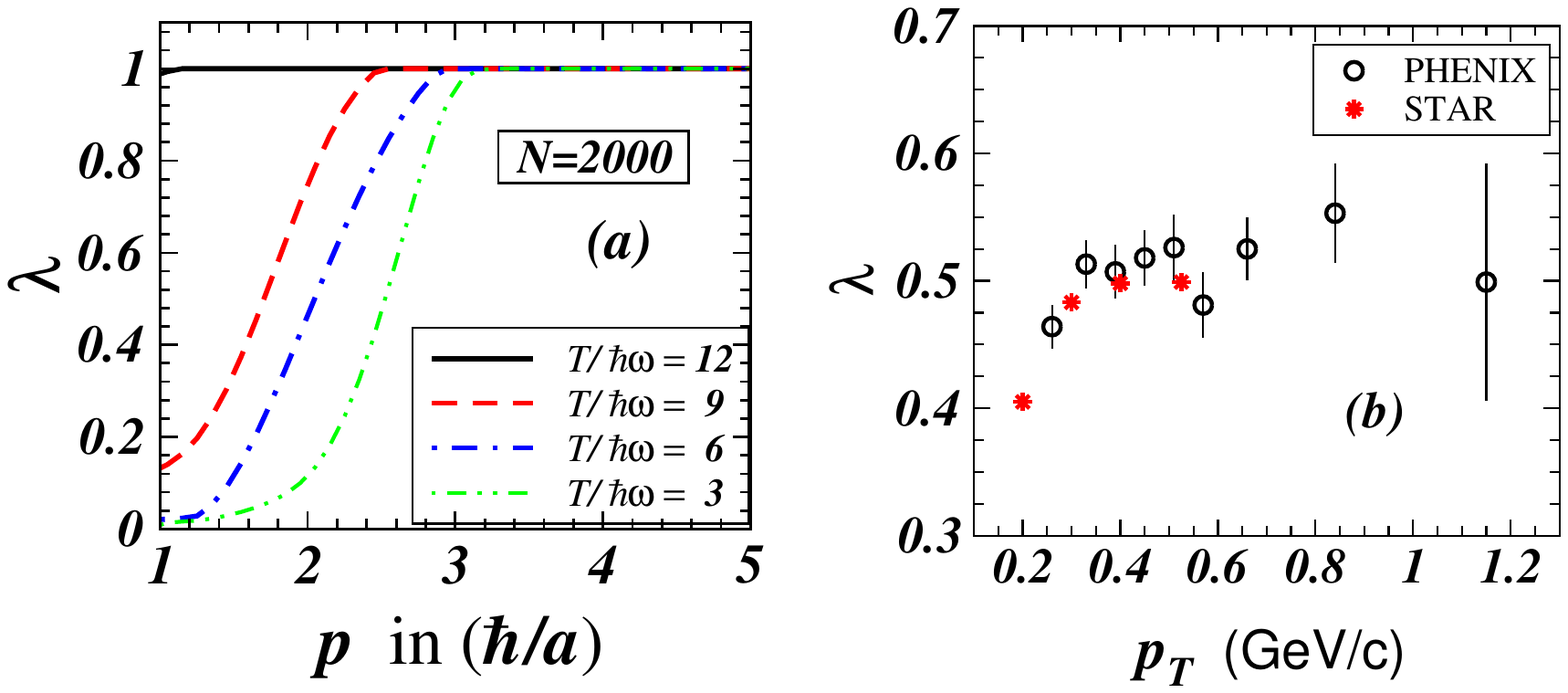}
\vspace*{-11.0cm}
\caption{(Color Online) (a) The parameter $\lambda$ as a function of
  $p$ for different temperatures for $N$=2000.  (b) Experimental
  measured values of $\lambda$ as a function of $p_T$ for AuAu
  Collisions at RHIC at $\sqrt{s_{\rm NN}}$=200 GeV from the PHENIX
  Collaboration \cite{PHE04} and the STAR Collaboration \cite{STA05}.}
\end{figure}
If one follows the standard phenomenological analysis and introduces
the ``chaoticity" parameter $\lambda$ to represent the intercept of
the correlation function at zero relative momentum, then this
parameter $\lambda$ is a function of the average pair momentum $p$ and
temperature $T$
\begin{eqnarray}
\lambda(p,T)=[C(p,q=0;T)-1].
\end{eqnarray}
We display explicitly the dependence $\lambda(p,T)$ as a function of
$p$ in Fig.\ 5(a) for different order parameters $T/\hbar \omega$, for
the case of $N=2000$.  At $T/ \hbar \omega$=12, which is above the
critical condensate order parameter of $T_c/ \hbar \omega$=10.97, the
$\lambda$ parameter is 1 for all $p$ values.  At $T/\hbar \omega$=9,
as $p$ increases the $\lambda$ parameter rises gradually from
$\sim$0.1 at $p = \hbar/a$ and reaches the constant value of 1 at
$p$=2.4$\hbar/a$.  At $T/\hbar \omega$=6 and 3, for which the systems
are significantly coherent with large condensate fractions, the
$\lambda$ parameter starts close to zero at $p$=$\hbar/a$, but as $p$
increases the $\lambda$ parameter increases gradually to unity at
$p$=2.9 and 3.1$\hbar/a$ for $T/\hbar \omega$=6 and 3 respectively.
The location where the $\lambda$ parameter attains unity changes with
temperature.  The lower the temperature, the greater is the value of
$p$ at which the $\lambda$ parameter attains unity.

We conclude from our results that the parameter $\lambda(p,T)$ is a
sensitive function of both $p$ and $T$, and $\lambda(p,T)=1$ is not a
consistent measure of the absence of the condensate fraction, as it
attains the value of unity in some kinematic regions for significantly
coherent systems with large condensate fractions at temperatures much
below $T_c$.  Only for the region of small $p$ will the parameter
$\lambda(p,T)$ be correlated with the chaotic
fraction $f_T(T)$ of the system.

It is interesting to note that experimentally measured values of
$\lambda$ from different collaborations and different method of
analysis \cite{PHE04,STA05} exhibit an increase of $\lambda$ as $p_T$
of the average momentum of the pair increase as shown in
Fig.\ 5(b). There is a similar trend of increasing $\lambda$ as a
function of $p_T$. This may be an indication of the dependence of the
correlation function on the average momentum of the pair arising for a
partially coherent pion source.  The increase of $\lambda$ as a
function of the average pair momentum has also been obtained in the
q-deformed harmonic oscillator model of Bose-Einstein correlations \cite{Gav06}.

\section{Bose-Einstein Condensation of Pions in their Men Fields
}

With regard to heavy-ion collisions at RHIC \& LHC, it is instructive
to raise the following question.  If we have a pion system that has a
root-mean squared radius $r_{\rm rms}$=10 fm, the number of identical
pions $N$ from a few hundred to a few thousand, at a freezeout
temperature $T$=80 to 160 MeV, typical of those revealed by
Bose-Einstein correlation measurements \cite{PHE04,STA05}, then, what
will be the condensate fraction $f_0$?  To answer this question, it is
useful to calculate the root-mean-squared radius $r_{\rm rms}$ of the
pion system as a function of the order parameter $T/\hbar \omega$ for
a pion system with $N$=250 to 2000 as shown in Fig.\ 6.  We can
schematically represent the functional relation between $r_{\rm rms} /
a$ and $T/\hbar \omega$ in Fig.\ 6 as
\begin{eqnarray}
r_{\rm rms} / a=f_N(T/\hbar\omega).
\end{eqnarray}
\begin{figure}[h]
\hspace*{2.0cm}
\includegraphics[scale=0.40]{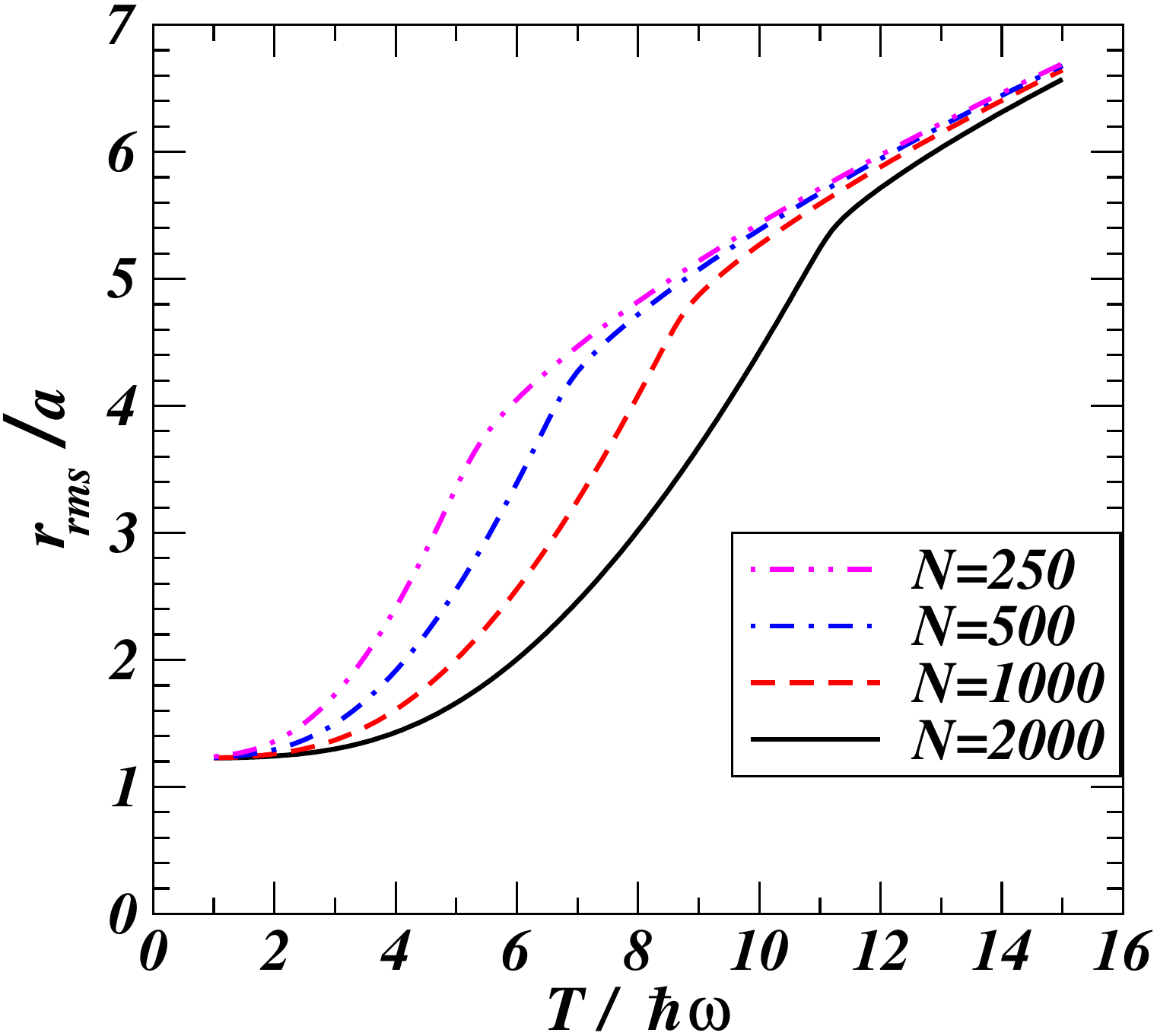}
\vspace*{-0.2cm}
\caption{(Color online) The root-mean-squared radius in unit of $a$
  and the root-mean-squared momentum in units of $\hbar/a$, as a
  function of $T/\hbar\omega$ for different numbers of identical
  bosons in the system.}
\vspace*{-0.3cm}
\end{figure}
\begin{figure}[h]
\hspace*{2.0cm}
\includegraphics[scale=0.40]{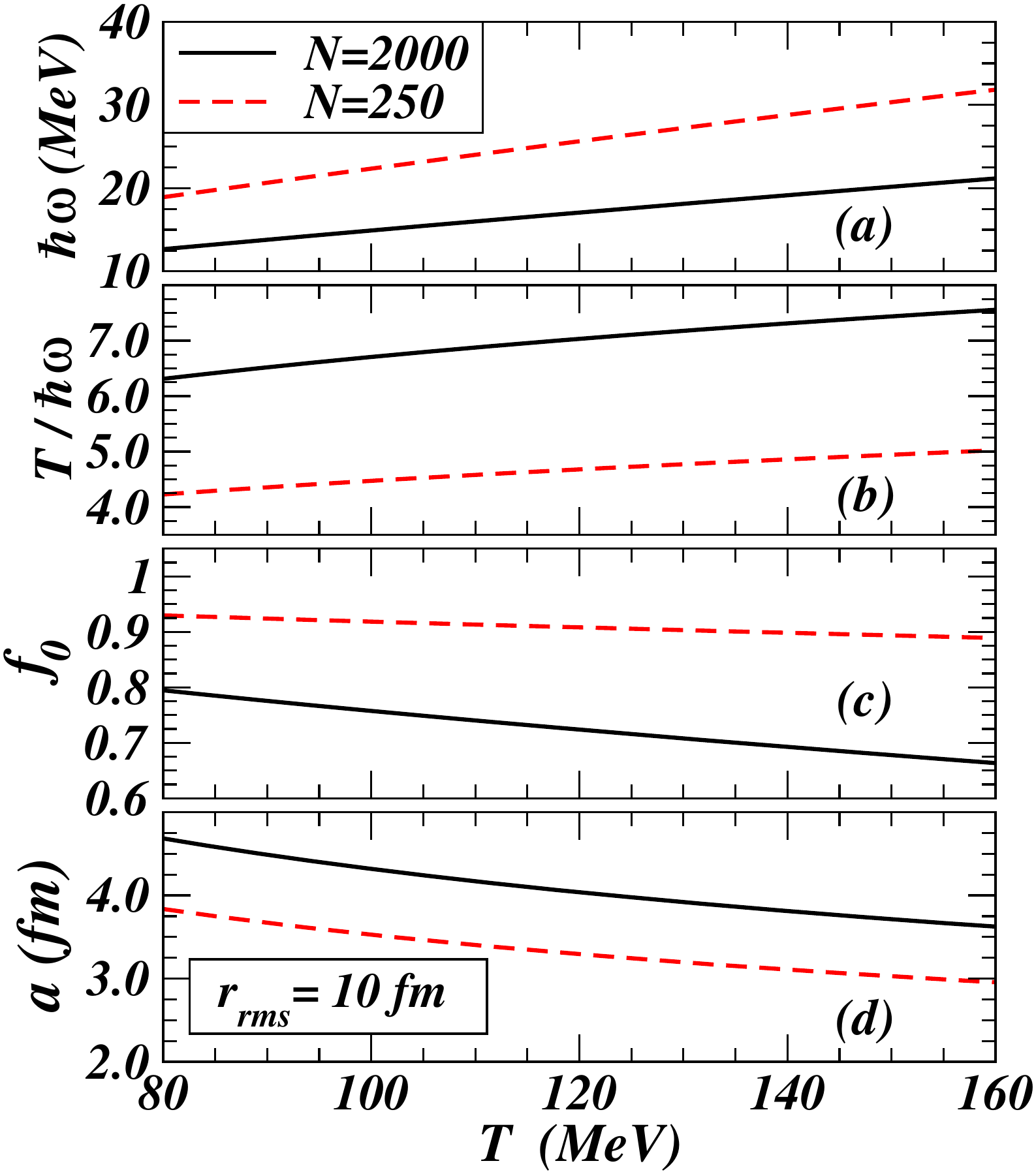}
\vspace*{-0.2cm}
\caption{(Color online) ($a$) the potential strength $\hbar \omega $,
  ($b$) the ratio $T/\hbar \omega$, ($c$) the condensate fraction
  $f_0$, and ($d$) the oscillator length parameter $a$ for
  non-relativistic boson systems with $N=2000$ and $N=250$ in a static
  equilibrium with a $r_{\rm rms}=10$ fm, plotted as a function of the
  temperature $T$.  }
\end{figure}
For a given value of $N$ and $r_{\rm rms}$, as $a$ is equal to
$\hbar/\sqrt{m_\pi \hbar \omega}$, the above equation contains only a
single variable $\hbar \omega$ that can be determined as a function of
$T$.  Subsequently, the order parameter $T/\hbar \omega$ and the
condensate fraction $f_0$ can also be determined as a function of $T$
as shown in Fig. 7.

One finds that for the pion system with a given
root-mean-squared radius of 10 fm, the value of $\hbar \omega $ ranges
from about 12 to 20 MeV for $N$=2000 and about 20 to 30 MeV for
$N$=250. The ratio of $T/\hbar\omega $ about 7 for $N$=2000, and is
about 4.5 for $N$=250, as shown in Fig. 7($b$).  From these ratios of
$T/\hbar\omega$, one can use Fig. 2 to find out the condensate
fraction.  The condensate fractions $f_0(T)$ for a pion gas at various
temperatures with $N$=2000 and $N$=250 are shown in Fig. 7($c$).  One
finds that $f_0(T)$ is about $0.67-0.8$ for $N$=2000 and is about
$0.9$ for $N$=250.

We reach the conclusion from the above study that if a non
relativistic pion system maintains a static equilibrium within its
mean field, and if it contains a root-mean-squared radius, a pion
number, and a temperature typical of those in high-energy heavy-ion
collisions, then it will contain a large fraction of the Bose-Einstein
pion condensate.  For a relativistic pion system, while the absolute
scale of the order parameter $T/\hbar \omega$ may change, the
condensate fraction $f_0$ remains substantial \cite{Won07}.  Pion
condensation will affect the parameter $\lambda$ in momentum
correlation measurements.

\section{Three-particle Correlations and Coherence}

Bose-Einstein condensation has important influence on the
three-particle correlation function.  We can determine the dependence
of the three-particle correlation function on the degree of
Bose-Einstein coherence in a way similar to what has been carried out
for two-particle correlations.

The extraction of the coherence properties from experimental
three-particle correlation data has the advantage that the problems of
the resonances can be minimized.  It has however the disadvantage that
the statistics in the number of three-particle events may be lowered
because of the restriction on the occurrence of three-particle
coincidences.

Recently there has much interest in three-particle correlation
measurements \cite{Abe14}.  Bose-Einstein condensation of pions in a
heavy-ion collision may suppress Bose-Einstein correlations.
Furthermore, initial conditions such as the color-glass condensate
(CGC) with the coherent production of partons \cite{McL94} may also
lead to condensate formation \cite{Bla12}.  Experimental results
indicate the presence of a substantial condensate fraction
\cite{Abe14}.  It is of interest to formulate an analytical model to
investigate how the three-particle correlation function will depend on
the coherence of the underlying boson system.
\begin{figure} [h]
\hspace*{1.2cm}
\includegraphics[scale=0.62]{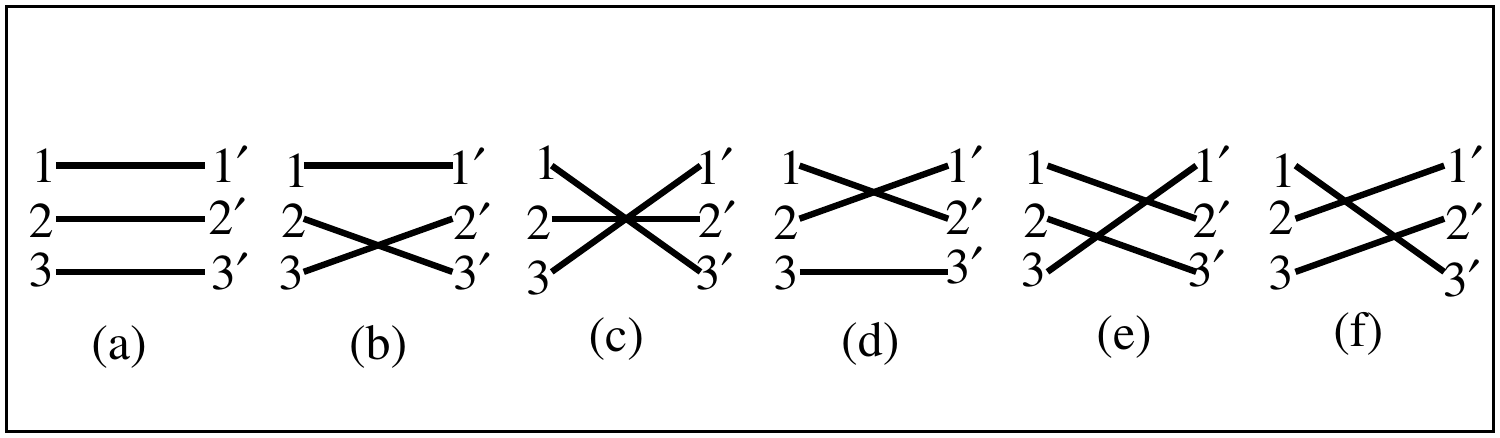}
\caption{Three-particle distribution function expanded in terms of
  one-particle distribution function in uncorrelated mean-field
  approximation. }
\end{figure}
In a completely chaotic source when multi-particle Bose-Einstein-type
correlations are neglected, the three-particle correlation function
can be written in terms of products of one-body distribution
functions:
\begin{eqnarray}
&&\hspace*{-0.5cm}G^{(3)}(p_1,p_2.p_3;p_1',p_2',p_3')
\nonumber\\
&&\hspace*{-0.5cm}= \! G^{(1)}(1,1') G^{(1)}(2,2')G^{(1)}(3,3') + G^{(1)}(1,2') G^{(1)}(2,1')G^{(1)}(3,3') 
\nonumber\\
&&\hspace*{-0.5cm}~+ G^{(1)}(1,3') G^{(1)}(2,2')G^{(1)}(3,1')+ G^{(1)}(1,1') G^{(1)}(2,3')G^{(1)}(3,2') 
\nonumber\\
&&\hspace*{-0.5cm} ~+ G^{(1)}(1,3') G^{(1)}(2,1')G^{(1)}(3,2')  + G^{(1)}(1,2') G^{(1)}(2,3')G^{(1)}(3,1'), 
\end{eqnarray}
as represented by the diagrams in Fig.\ 8.  With Bose-Einstein
correlations, we can generalize our two-particle correlation case to
the three-particle correlation functions and write down the
three-particle correlation function as
\begin{eqnarray}
C(p_1,p_2,p_3)\!\!\!\!\!\!\!\!&&\equiv 
 \frac{G^{(3)}(1,2,3;1',2',3')}{G^{(1)}(1,1') G^{(1)}(2,2')G^{(1)}(3,3')}\biggr |_{1' \to 1, 2' \to 2, 3' \to 3}
\nonumber\\
&& =1+\frac{G^{(1)}(1,2) G^{(1)}(2,1) - N_0u_0^2(p_1) u_0^2(p_2)   }{G^{(1)}(1,1) G^{(1)}(2,2)}
\nonumber\\
&&~~~~~+\frac{G^{(1)}(1,3) G^{(1)}(3,1)-N_0^2 u_0^2(p_1) u_0^2(p_3) }{G^{(1)}(1,1) G^{(1)}(3,3)}
\nonumber\\
&&~~~~~+ \frac{G^{(1)}(2,3) G^{(1)}(3,2)-N_0^2 u_0^2(p_2) u_0^2(p_3) }{G^{(1)}(2,2) G^{(1)}(3,3)}
\nonumber\\
&&
+ \frac{G^{(1)}(1,3) G^{(1)}(2,1)G^{(1)}(3,2)  
-N_0^3  u_0^2(p_1)  u_0^2(p_2) u_0^2(p_3)
}{G^{(1)}(1,1) G^{(1)}(2,2)G^{(1)}(3,3)}
\nonumber\\
&&
+ \frac{ G^{(1)}(1,2) G^{(1)}(2,3)G^{(1)}(3,1)
-N_0^3  u_0^2(p_1)  u_0^2(p_2) u_0^2(p_3)
}{G^{(1)}(1,1) G^{(1)}(2,2)G^{(1)}(3,3)}.
\label{eq3p}
\end{eqnarray}
The above correlation function $C(\bb p_1, \bb p_2, \bb p_3)$
possesses the proper coherent and chaotic limits.  For a nearly
completely coherent source with almost all particles populating the
ground condensate state, $N_0$$\to$$ N$, the terms in the numerator
cancel each other and we have $C(\bb p_1, \bb p_2, \bb p_3)$=1, and
the BE correlation is absent.  For the other extreme of a completely
chaotic source with $N_0$$\ll$$N$, the second terms in the numerators
proportional to $N_0^2$ give negligible contribution and can be
neglected.  The correlation function $C(\bb p_1, \bb p_2, \bb p_3)$
becomes the usual BE correlation for a completely chaotic source.
These results will allow the evaluation of the three-particle
correlation function using the functions of $G^{(1)}(\bb p_1, \bb p_2)
$ and $u_0^2(\bb p)$ in Eqs. (\ref{eq22})-(\ref{eq23A}).  Different
ways of re-combining some of the terms in Eq.\ (\ref{eq3p}) in terms
of two-particle correlation functions may allow one to extract
quantities that minimize the systematic errors in two-particle
correlation function measurements.

\section{Conclusions and Summary}

A proper framework to study Bose-Einstein correlations is the theory
of Bose-Einstein condensation.  We examine the condition for the
occurrence of the Bose-Einstein condensation in an exactly solvable
model. We place identical bosons in a spherical harmonic oscillator
potential that arises either externally or approximately from its own
mean fields.  The order parameter is $T/\hbar\omega$, the ratio of the
temperature to the energy gap between the lowest and the first excited
single-particle state.  The degree of chaoticity or condensation is
quantified by the condensate fraction $f_0=N_0/N$ which specifies the
transition from a chaotic state to a coherent condensate state.  The
condensate fraction $f_0$ is a cubic function of the order parameter
$T/\hbar \omega$.  The critical order parameter $T_c/\hbar \omega$
varies with the boson number $N$ as $T_c/\hbar \omega$=$
0.6777N^{0.3666}$.  The transition from the completely chaotic state
with $f_0$=0 to the completely coherent state with $f_0$$ \to$1 is a
second-order-type transition under a gradual decrease of the order
parameter $T/\hbar \omega$.  It is not a first-order phase transition.
A pion gas with $r_{\rm rms}, T$, and $N$, typical of those in RHIC
and LHC, is expected to contain a large condensate fraction and a high
degree of suppression of Bose-Einstein correlation.

The evaluation of the two-particle correlation function indicates that
the usual ``chaoticity parameter" $\lambda$ can only be interpreted as
an experimental tool to label the intercept of the correlation
function $C(\bb p, \bb q)$ at $\bb q$=0.  The parameter $\lambda$ is
correlated with the degree of chaoticity only for small values of $p$
but is at variance from such an interpretation of chaoticity at high
values of $p$, as shown in Figs.\ 4 and 5(a).

We have written out the functional form of the three-particle
distribution function as a function of the momenta of the three
particles that contains the proper chaotic and coherent limits.  It
permits the description for the transition from the chaotic states to
coherent states. These results will allow the evaluation of the
three-particle correlation function in an exactly solvable problem
that will assist the comparison with three-particle correlation
measurements.

\vspace*{0.6cm}
\noindent{\Large \bf Acknowledgement}

\vspace*{0.6cm} This work was supported in part by the Division of
Nuclear Physics, U.S. Department of Energy, Contract
No. DE-AC05-00OR22725, and the National Natural Science Foundation of
China, Contract Nos. 11075027 and 11275037.


\begin{thebibliography}{10}

\bibitem{Han56}
R. Hanbury~Brown and R.~Q. Twiss, Nature {\bf 178},  1046  (1956).
\
\bibitem{Gla63}
R. J. Glauber, Phys . Rev. Lett. {\bf 10}, 84 (1963); \\
R. J. Glauber, Phys. Rev. {\bf 130}, 2529 (1963); \\
R. J. Glauber, Phys. Rev. {\bf 130}, 2766 (1963).

\bibitem{Gyu79}
M. Gyulassy, K. K. Kauffman, L. W. Wilson, Phys. Rev. {\bf C20},  2267 (1979).

\bibitem{Vos94}
D.N. Voskresensky, J. Exp. Theor. Phys. {\bf 78},  793  (1994); 
E.E. Kolomeitsev, D.N. Voskresensky, Phys. Atom. Nucl. {\bf 58},  2082 (1995);
E.E. Kolomeitsev, Burkhard Kampfer, D.N. Voskresensky, Acta Phys. Polon. {\bf B27},  3263 (1996); \\
D.N. Voskresensky, Phys.Atom.Nucl. {\bf 59},  2015 (1996). 

\bibitem{Wie99}
U. Wiedemann and U. Heinz, Phys. Rep. {\bf 319}, 145  (1999).

\bibitem{Wei00}
R. M. Weiner,  Phys. Rept. {\bf 327}  249  (2000).

\bibitem{Lis05}
M. A. Lisa, S. Pratt, R. Soltz, and U. Wiedemann, Ann. Rev. Nucl. Part. Sci. {\bf 55},
 357 (2005).

\bibitem{Won94}
For a pedagogical discussion, see Chapter XVII of
  C. Y. Wong, {\it Introduction to High-Energy Heavy-Ion
  Collisions}, World Scientific Publisher, 1994.

\bibitem{Pol96}
H. D. Politzer,
Phys. Rev.  {\bf A54}, 5048 (1996).

\bibitem{Nar99}
M. Naraschewski and R. Glauber, Phys. Rev.  {\bf A59}, 4595 (1999).

\bibitem{Gom06} 
J. Viana Gomes, A. Perrin, M. Schellekens, D. Boiron,
C. I. Westbrook, and Michael Belsley,  Rev. A {\bf 74}, 053607
(2006);
 M. Yasuda and F. Shimizu, Phys. Rev. Lett. {\bf 77}, 3090 (1996);
D. Hellweg, L. Cacciapuoti, M. Kottke, T. Schulte, K. Sengstock, 
W. Ertmer, and J. J. Arlt, Phys. Rev. Lett. 91, 010406 (2003);
M. Greiner, C. A. Regal, J. T. Stewart, and D. S. Jin, Phys. Rev. Lett. {\bf 94}, 110401 (2005);
S. F\" olling, F. Gerbier, A. Widera, O. Mandel, T. Gericke, 
and I. Bloch, Nature {\bf 434}, 481 (2005);
A. Ottl, S. Ritter, M. Kohl, and T. Esslinger, 
Phys. Rev. Lett. {\bf 95}, 090404 (2005);
M. Schellekens, R. Hoppeler, A. Perrin, J. Viana
Gomes, D. Boiron, A. Aspect, and C. I. Westbrook, Science {\bf 310}, 648
(2005);
J. Esteve, J.-B. Trebbia, T. Schumm, A. Aspect,
C. I. Westbrook, and I. Bouchoule, Phys. Rev. Lett. {\bf 96}, 130403 (2006).

\bibitem{Fow77}
C. N. Fowler and R. M. Weiner, Phys. Lett. {\bf B70},  201 (1977);\\
C. N. Fowler and R. M. Weiner, Phys. Rev.  {\bf D17},   3118 (1978);\\
C. N. Fowler, N. Stelte, and R. M. Weiner, Nucl. Phys. {\bf A319},  349  (1979).

\bibitem{Pra93}
S. Pratt, Phys. Lett.  {\bf B301},  159 (1993).

\bibitem{CsoZim98}
T. Cs\"{o}rg\H{o} and J. Zim\'{a}nyi, Phys. Rev. Lett. {\bf 80}  916 (1998);\\
J. Zim\'{a}nyi and T. Cs\"{o}rg\H{o}, Heavy Ion Phys. {\bf 9}
 241 (1999).

\bibitem{Won07}
C. Y. Wong and W. N. Zhang, Phys. Rev {\bf C76}, 034905 (2007).

\bibitem{Liu13}
J. Liu, P. Ru, and W. N. Zhang, Int. J. Mod. Phys.  {\bf E22},   1350083 (2013).

\bibitem{LiuZha14}
J. Liu, P. Ru, and W. N. Zhang, C. Y. Wong, Jour. Phys. {\bf G41}, 125101 (2014).

\bibitem{Gav06}
A. M. Gavrilik,
Symmetry, Integrability and Geometry, {\bf 2}, 1 (2006);\\
A. M. Gavrilik, SIGMA, {\bf 2}, 74  (2006) [hep-ph/0512357];\\
A. M. Gavrilik, A. Rebesh, Mod. Phys. Lett.  {\bf B25}, 1150030 (2012);\\
A. M. Gavrilik, I. Kachurik, A. Rebesh, [arXiv:1309.1363];\\
A. M. Gavrilik, I. Kachurik, Y. Mishchenko, J. Phys.  {\bf A56}, 948 (2011);\\
A. M. Gavrilik, Y. Mishchenko,  Phys. Lett. {\bf A376}, 1596 (2012);\\
A. M. Gavrilik, Y. Mishchenko, Ukr. J. Phys. {\bf 58}, 1171 (2013).

\bibitem{Sin13}
Yu. M. Sinyukov and V. M. Shapoval,
Phys. Rev.  {\bf D87}, 094024 (2013).

\bibitem{McL94}
L.McLerran and R.Venugopalan, Phys.  Rev. {\bf D49}, 2233 (1994);\\
L.McLerran and R.Venugopalan, Phys. Rev. {\bf D49}, 3352 (1994).

\bibitem{Bla12}
J. P. Blaizot et al., Nucl. Phys. {\bf A873}, 68 (2012).

\bibitem{Abe14}
B. Abelev et al. (ALICE Collaboration), Phys. Rev. {\bf C89}, 024911 (2014). 

\bibitem{NA44}
H. Boggild et al. (NA44 Collaboration), Phys. Lett.  {\bf B455} 77  (1999);\\
I. G. Bearden et al. (NA44 Collaboration), Phys. Lett. {\bf B517} 25  (2001).

\bibitem{WA98}
M. M. Aggarwal et al. (WA98 Collaboration), Phys. Rev. Lett. {\bf 85}  2895 (2000).

\bibitem{STAR}
J. Adams et al. (STAR Collaboration), Phys. Rev. Lett. {\bf 91}  262301 (2001).

\bibitem{Mor05}
K. Morita, S. Muroya, and H. Nakamura,  Prog. Theo. Phys. 
{\bf 114}, 583 (2005).


\bibitem{Zaj87}
W. A. Zajc, Phys. Rev. {\bf D35}, 3396 (1987).

\bibitem{Biy90}
 M. Biyajima, A. Bartl, T. Mizoguchi, N. Suzuki and O. Terazawa, Prog. Theor. Phys. {\bf 84} 931
(1990).

\bibitem{And91}
I. V. Andreev, M. Pl\" umer,  R. M. Weiner, Phys . Rev. Lett. {\bf 67}, 3475 (1991);\\
I. V. Andreev, M. Pl\"umer,  R. M. Weiner, Int. J. Mod.
Phys.{\bf  A8}, 4577  (1993).

\bibitem{Zha93}
W. Z. Zhang, Y. M. Liu, S. Wang $et~ al.$, PHys. Rev. {\bf C47}, 795 (1993);\\
W. N. Zhang, Y. M. Liu, L. Huo  $et~ al.$, Phys. Rev.  {\bf C51}, 922  (1995) ; \\
W. N. Zhang, L. Huo, X. J. Chen  $et~ al.$, Phys. Rev. {\bf C58},  2311 (1998); \\
W. N. Zhang, G. X. Tang, X. J. Chen  $et~ al.$, Phys. Rev. {\bf C62},   044903 (2000). 

\bibitem{Cha95}
W. Q. Chao, C. S. Gao, and Q. H. Zhang,  J.  Phys.  {\bf G21}, 847
 (1995);\\ 
Q. H. Zhang, W. Q. Chao, and C. S. Gao, Phys. Rev. {\bf C52},
2064  (1995).

\bibitem{Hei97}
U. Heinz and Q. H. Zhang, Phys. Rev. {\bf C56}, 426 (1997); \\ U. Heinz and A. Sugarbaker, Phys. Rev. {\bf C70}, 054908 (2004).

\bibitem{Nak99}
 H. Nakamura and R. Seki, Phys. Rev.  {\bf C60} , 064904(1999);\\
 H. Nakamura and R. Seki, Phys. Rev. {\bf C61},  054905 (2000).

\bibitem{Cso02}
T. Cs\"org\"o, Heavy Ion Phys. {\bf 15}, 1 (2002).

\bibitem{Gla06}
R. Glauber, Nucl. Phys. {\bf A774}, 3 (2006).

\bibitem{Gla59} R. J. Glauber, in Lectures in Theoretical Physics,
  edited by W. E. Brittin and L. G. Dunham (Interscience, New York,
  1959), Vol 1, p. 315.

\bibitem{PHE04}
S. S. Adler $et~al.$, (PHENIX Collaboration), Phys. Rev. Lett. {\bf 93}, 152302 (2004).

\bibitem{STA05}
J. Adams $et~al.$, (STAR Collaboration), Phys. Rev.  {\bf C71}, 044906 (2005).

\end{thebibliography}
\end{document}